\newif\ifproofs

\newif\ifFullVersion

\documentclass[9pt,final,twocolumn]{IEEEtran}
\IEEEoverridecommandlockouts
\usepackage{balance}
\usepackage{amsmath}
\usepackage{amsthm}
\usepackage[bookmarks,colorlinks]{hyperref}
\usepackage{cite}
\usepackage{cases}
\usepackage{amssymb}
\usepackage{algorithm}
\usepackage{algorithmic}
\usepackage{graphicx}
\usepackage{subcaption}
\usepackage{color}
\usepackage{enumerate}
\usepackage{xcolor}
\usepackage{acronym}

\renewcommand{\algorithmicrequire}{ \textbf{Initialize:}} 
\renewcommand{\algorithmicensure}{ \textbf{Iteration:}}

\acrodef{adc}[ADC]{analog-to-digital convertor}
\acrodef{bilispa}[BiLiSPA]{bit-limited sparse linear array}
\acrodef{dft}[DFT]{discrete Fourier transform}
\acrodef{cs}[CS]{compressed sensing}
\acrodef{sla}[SLA]{sparse linear array}
\acrodef{ula}[ULA]{uniform linear array}
\acrodef{doa}[DOA]{direction of arrival}
\acrodef{had}[HAD]{hybrid analog and digital}
\acrodef{mse}[MSE]{mean square error}
\acrodef{sdp}[SDP]{semi-definite programming}
\acrodef{mimo}[MIMO]{multiple input and multiple output}

\newcommand{\figWidth}{2.5in}
\newcommand{\figHeight}{1.75in}

\DeclareMathOperator{\diag}{diag}

\title{\huge LiQuiD-MIMO Radar: Distributed MIMO Radar with Low-Bit Quantization}
  
\author{Yikun~Xiang, Feng~Xi, Shengyao~Chen
\thanks{Y. Xiang, F. Xi, and S. Chen are with the Department
of Electronic Engineering, Nanjing University of Science and Technology, Nanjing 210094 China (email: xifeng@njust.edu.cn). 
This work  was supported in part by the Natural Science Foundation of Jiangsu Province, China, under grant No.BK20221486. }
\vspace{-1.0cm}
}
\begin{document}
%
\maketitle
\pagestyle{empty}  
\thispagestyle{empty} 
\begin{abstract}
 Distributed MIMO radar is known to achieve superior sensing performance by employing widely separated antennas.
 However, it is challenging to implement a low-complexity distributed MIMO radar due to the complex operations at both the receivers and the fusion center.
 This work proposed a low-bit quantized distributed MIMO (LiQuiD-MIMO) radar to significantly reduce the burden of signal acquisition and data transmission.
 In the LiQuiD-MIMO radar, the widely-separated receivers are restricted to operating with low-resolution ADCs and deliver the low-bit quantized data to the fusion center. At the fusion center, the induced quantization distortion is explicitly compensated via digital processing. By exploiting the inherent structure of our problem, a quantized version of the robust principal component analysis (RPCA) problem is formulated to simultaneously recover the low-rank target information matrices as well as the sparse data transmission errors.
 The least squares-based method is then employed to estimate the targets' positions and velocities from the recovered target information matrices.
 Numerical experiments demonstrate that the proposed LiQuiD-MIMO radar, configured with the developed algorithm, can achieve accurate target parameter estimation.
 
\end{abstract}
\vspace{-0.4cm}
\section{Introduction}
\label{sec:intro}
\vspace{-0.1cm}
Over the last few years, multiple-input-multiple-output (MIMO) radar \cite{Fishler2004,Li2007,Haimovich2008} has attracted significant attention in civil and military applications, due to its superior performance for target detection and localization.
Depending on the antenna configurations, two typical MIMO radar architectures are developed, namely collocated MIMO radar \cite{Li2007} and distributed MIMO radar \cite{Haimovich2008}.
The antennas in collocated MIMO radar are closely spaced while those are widely separated in distributed MIMO radar.
In this paper, we mainly focus on distributed MIMO radar.

In distributed MIMO radar \cite{Haimovich2008}, the spatial diversity is exploited to improve the target detection performance, since each pair of transmit (Tx) and receive (Rx) antennas observes a target from different angles.
The main challenge of distributed MIMO radar is to jointly process the signals received by those widely separated antennas.
Typically, in a distributed MIMO radar, each receiver is equipped with limited hardware for local processing.
To jointly process all the data collected by each receiver, the results of local processing have to be delivered to a fusion center to make the final decision.
However, with the widespread application of large-scale arrays, the complexity of such a MIMO radar system and the corresponding data transmission and processing becomes prohibitively high.
The aim of this paper is to develop a novel architecture to cope with the system complexity, data transmission volume, and energy consumption of such complicated systems.

The leading approach in the literature to reduce the complexity of a MIMO radar system is to utilize compressive sensing \cite{eldar2012compressed} to reduce the sampling rate \cite{Cohen2018TSP, Mishra2020TAES}, the number of antennas \cite{Rossi2014TSP}, and the amount of samples \cite{Yu2010JSTSP, SunTAES2015}.
These works mainly focus on the sampling aspect of signal acquisition and ignore the effect of quantization, assuming the use of high-resolution analog-to-digital converters (ADCs).
Unfortunately, the energy consumption and cost of an ADC grow exponentially with the bit depth \cite{Walden1999}.

Recent years have witnessed a growing interest in signal acquisition operating with low-resolution ADCs.
The major issue in low-bit quantization is developing signal processing methodologies to compensate for the distortion induced in quantization.
A variety of signal acquisition architectures as well as signal processing algorithms built upon low-resolution ADCs have been proposed in a multitude of different applications, including radars \cite{Zahabi2017ICASSP,Xi2018SAM,Ameri2019TSP,Mazher2021}, communications \cite{Wan2020TWC,Myers2020ICASSP}, joint radar-communication systems \cite{Kumari2020ICASSP,Zhu2021ICASSP}, array signal processing \cite{Liu2017ICASSP,Sedighi2021SPL,Sedighi2021TSP}, and so on.
In the context of MIMO radars, the works \cite{Xi2020TSP,Xi2020TAES,Xi2021TSP} proposed to use one-bit or low-bit quantization to reduce the system complexity and modified the signal processing to account for low-bit quantized observations.
However, these works focus on collocated MIMO radar, which are not suitable for distributed MIMO radar due to the totally different architectures.

In this paper, we propose a new distributed MIMO radar architecture, referred to as the Low-bit Quantized Distributed MIMO (LiQuiD-MIMO) radar, in which each receiver quantizes the received signal via low-resolution ADCs, and then transmits the low-bit quantized data to the fusion center for sophisticated processing.
Owing to the low-bit quantization, the cost, energy consumption, and hardware complexity at each receiver are largely reduced, and the amount of data transmitted from the receiver located remotely to the fusion center is greatly decreased.
By exploiting the low-rank structure of the target information matrix (TIM) and the sparsity of the data transmission error (DTE), we formulate the target estimation problem at the fusion center as a quantized version of robust principal component analysis (RPCA) problem \cite{Candes2011}, referred to as QRPCA problem.
To deal with the QRPCA problem, an accelerated proximal gradient (APG)-based algorithm is proposed to iteratively recover the TIM and DTE matrices.
Finally, the least squares (LS)-based method is applied to combine all the information collected by each pair of Tx-Rx antennas and output the position and velocity estimation of the targets.

The rest of the paper is organized as follows: Section \ref{sec:model} introduces the LiQuiD-MIMO radar system. 
Section \ref{sec:method} formulates the QRPCA problem and develops the APG-based algorithm to solve it.
The target parameter estimation from the solutions to the QRPCA problem is detailed in Section \ref{sec:parameter_estimation}.
Sections \ref{sec:Sims} and \ref{sec:conclusion} provide numerical simulations and concluding remarks, respectively.


\vspace{-0.2cm}
\section{LiQuiD-MIMO Radar Model}
\label{sec:model}
\vspace{-0.1cm}
In this section, we first describe the signal model for our distributed MIMO radar system and then introduce the sampling and quantization scheme with low-resolution ADCs in the LiQuiD-MIMO radar.

\vspace{-0.2cm}
\subsection{Signal Model}
\label{subsec:signal_model}
\vspace{-0.1cm}
Consider a distributed MIMO radar system equipped with $M_t$ transmit antennas and $M_r$ receive antennas, which are widely separated in a two-dimensional (2-D) Cartesian coordinate system. 
The positions of the $m$th transmitter antenna and the $n$th receiver antenna are denoted as $\mathbf{p}_t^{(m)}=[x_t^{(m)}, y_t^{(m)}]^T$ and $\mathbf{p}_r^{(n)}=[x_r^{(n)}, y_r^{(n)}]^T$, respectively. 
A set of $M_t$ narrowband and orthogonal waveforms, denoted as $s_1(t), \cdots, s_{M_t}(t)$, are transmitted in pulses, with a pulse repetition interval (PRI) $T_{\text{PRI}}$. 
We assume that each coherent processing interval (CPI) includes $Q$ pulses, i.e., the $m$th transmitted signal during one CPI is $\sum_{q=0}^{Q-1} s_m(t-qT_{\text{PRI}})$.

To guarantee the orthogonality of the transmitted waveforms, we consider orthogonality achieved using frequency division multiple access (FDMA) signaling. 
In FDMA, the carrier frequency of the $m$th transmitted signal is $f_m = f_0 + (m-1)\Delta f$, $m=1,\cdots, M_t$, where $f_0$ is the reference carrier frequency and $\Delta f$ is the frequency increment across transmit antennas, satisfying that $\Delta f \geq B_0$ with $B_0$ being the bandwidth of $s_m(t)$.

We assume that all the targets are distributed in the same 2-D plane where the transmit and receive antennas are located. 
Without loss of generality, we can extend the analysis in this paper to the three-dimensional (3-D) case.
Suppose there are a total of $K$ targets located in the radar's unambiguous time-frequency region. 
The $k$th target is located at the position $\mathbf{p}^{(k)}=[x_k, y_k]^T$, moving at a velocity of $\mathbf{v}^{(k)}=[v_x^{(k)}, v_y^{(k)}]^T$. 
In the distributed MIMO radar, each of the $M_t$ transmitted signals are reflected by the targets and collected by each of the $M_r$ receive antennas.
For the $k$th target, the radar echoes from the $m$th transmit antenna to the $n$th receive antenna are characterized by the following parameters: its reflection coefficient $\tilde{\beta}_{mn}^{(k)}$, the time delay $\tau_{mn}^{(k)}$, 
and the Doppler frequency $f_{mn}^{(k)}$. We assume that these parameters are constant during one CPI.
Accodding to the geometric relationship, $\tau_{mn}^{(k)}$ and $f_{mn}^{(k)}$ are, respectively, determined by the following equations:
\begin{equation}
    \tau_{mn}^{(k)}=\frac{ \Vert \textbf{p}^{(k)} - \textbf{p}_t^{(k)} \Vert 
    + \Vert \textbf{p}^{(k)} - \textbf{p}_r^{(k)} \Vert }{c},
\end{equation}
\begin{equation}\label{eqn:f_mn}
    f_{mn}^{(k)} = \frac{f_m}{c} \left( 
    \frac{ \langle \textbf{v}^{(k)}, \textbf{p}^{(k)}-\textbf{p}_t^{(m)} \rangle }{ \Vert \textbf{p}^{(k)}-\textbf{p}_t^{(m)} \Vert } 
    + \frac{ \langle \textbf{v}^{(k)}, \textbf{p}^{(k)}-\textbf{p}_r^{(m)} \rangle }{ \Vert \textbf{p}^{(k)}-\textbf{p}_r^{(m)} \Vert }
    \right),
\end{equation}
where $c$ is the speed of light.

Then the received signal at the $n$th receive antenna can be expressed as
\begin{equation}
    \begin{split}
        y_n(t) = \sum_{q=0}^{Q-1}\sum_{m=1}^{M_t}\sum_{k=1}^{K}\tilde{\beta}_{mn}^{(k)}
        s_m(t-\tau_{mn}^{(k)}-qT_{\rm{PRI}}) \\
        \times e^{ j2\pi(f_m+f_{mn}^{(k)})(t-\tau_{mn}^{(k)}) } + w_n(t)
    \end{split}
\end{equation}
where $w_n(t)$ is the additive spatio-temporally white, zero mean Gaussian noise with variance $\sigma_n^2$. After demodulation and channel separation, the received signal at the $n$th receiver from the $m$th transmit antenna is 
\begin{equation}\label{eqn:ymn_t}
    \begin{split}
        y_{mn}(t) = \sum_{q=0}^{Q-1}\sum_{k=1}^{K} \beta_{mn}^{(k)}
        s_m(t-\tau_{mn}^{(k)}-qT_{\rm{PRI}}) \\
        \times e^{ j2\pi f_{mn}^{(k)}qT_{\rm{PRI}}} + w_{mn}(t)
    \end{split}
\end{equation}
where $\beta_{mn}^{(k)}=\tilde{\beta}_{mn}^{(k)}e^{ -j2\pi(f_m+f_{mn}^{(k)})\tau_{mn}^{(k)} }$, and $w_{mn}(t)$ denotes the noise in the corresponding transmission channel. In the above, we use the approximation $e^{ j2\pi f_{mn}^{(k)}qt}\approx e^{ j2\pi f_{mn}^{(k)}qT_{\rm{PRI}}}$ which follows the assumption that the targets are slow moving and have low acceleration.

\vspace{-0.2cm}
\subsection{Sampling and Quantization with Low-Resolution ADCs}
\label{subsec:samp_quan}
\vspace{-0.1cm}

In classic distributed MIMO radar, the received signal $y_{mn}(t)$ is sampled at the Nyquist rate and quantized by high-resolution ADCs, such that the quantization distortion is ignored in the subsequential signal processing.
Let $T_s$ be the Nyquist sampling interval. 
For each pulse received, we collect $L_{mn}$ unambiguous samples, where $L_{mn} = \lfloor\frac{T_p + \tau_{mn}^{\text{(max)}}}{T_s}\rfloor$ with $T_p$ and $\tau_{mn}^{\text{(max)}}$ denoting the pulse duration and maximum possible time-delay of targets, respectively.
For simplicity, we assume $T_p = NT_s$.
Then the samples of the time delayed signal $s_m(t-\tau_{mn}^{(k)})$ can be equivalently represented as $\mathbf{C}_{L_{mn}^{(k)}}^T\mathbf{s}_m$, where $\mathbf{C}_{L_{mn}^{(k)}}= \left[\mathbf{0}_{N\times L_{mn}^{(k)}} \quad \mathbf{I}_N \quad\mathbf{0}_{N\times (L_{mn}-N-L_{mn}^{(k)})}\right] \in\mathbb{C}^{N\times L_{mn}}$ with $L_{mn}^{(k)}=\lfloor\frac{\tau_{mn}^{(k)}}{T_s}\rfloor$, and $\mathbf{s}_m\in\mathbb{C}^N$ is the sampled $m$th transmit waveform.

After the Nyquist sampling and high-resolution quantization, the data collected during the $q$th pulse can be expressed as
\begin{equation}
\mathbf{y}_{mn}^{(q)}=\mathbf{x}_{mn}^{(q)}+\mathbf{w}_{mn}^{(q)}, \quad q=0,...,Q-1,
\end{equation}
where $\mathbf{x}_{mn}^{(q)}, \mathbf{w}_{mn}^{(q)}\in\mathbb{C}^{L_{mn}}$ are the the sampled signal vector and noise vector, respectively.
According to (\ref{eqn:ymn_t}), $\mathbf{x}_{mn}^{(q)}$ can be equivalently represented as
\begin{equation}
  \mathbf{x}_{mn}^{(q)}=\sum_{k=1}^{K} \beta_{mn}^{(k)}e^{j2\pi f_{mn}^{(k)}qT_{\rm{PRI}}}\mathbf{C}_{L_{mn}^{(k)}}^T \mathbf{s}_{m} = \mathbf{A}_{mn}\boldsymbol{\Lambda}_{mn}\mathbf{b}_{mn}^{(q)},
\end{equation}
where $\mathbf{A}_{mn} = \left[\mathbf{C}_{L_{mn}^{(1)}}^T\mathbf{s}_m,...,\mathbf{C}_{L_{mn}^{(K)}}^T\mathbf{s}_m \right] \in \mathbb{C}^{L_{mn} \times K}$, $\mathbf{b}_{mn}^{(q)}= \left[e^{ j2\pi f_{mn}^{(1)}qT_{\rm{PRI}} },\cdots,e^{ j2\pi f_{mn}^{(K)}qT_{\rm{PRI}} } \right]^T \in\mathbb{C}^K$, and
$\boldsymbol{\Lambda}_{mn} = \diag\{[\beta_{mn}^{(1)},\cdots,\beta_{mn}^{(K)}]\}$ is a $K\times K$ diagonal matrix.

Stacking the data $\{\mathbf{y}_{mn}^{(q)}\}_{q=0}^{Q-1}$ of all the $Q$ pulses into a $L_{mn}\times Q$ matrix, we have
\begin{equation}\label{eqn:Y_mn}
    \mathbf{Y}_{mn}=\mathbf{X}_{mn}+\mathbf{W}_{mn},
\end{equation}
where $\mathbf{X}_{mn}= \mathbf{A}_{mn}\boldsymbol{\Lambda}_{mn}\mathbf{B}_{mn}$ with $\mathbf{B}_{mn} = [\mathbf{b}_{mn}^{(0)}, \cdots, \mathbf{b}_{mn}^{(Q-1)}]$.
Since $\mathbf{X}_{mn}$ is determined by the unknown target parameters, we refer to it as the target information matrix (TIM).

Finally, the $n$th receive antenna will send the data set $\{\mathbf{Y}_{mn}\}_{m=1}^{M_t}$ to the fusion center for target parameter estimation. 
Thus, there are totally $M_tM_r$ raw data matrices being sent, i.e., the amount of data is $M_tM_rL_{mn}Q\tilde{b}$ if each data is quantized into $\tilde{b}$ bits.
In classic distributed MIMO radar, to avoid the effect of quantization distortion, the typical value of $\tilde{b}$ is $10\sim 16$, leading to a large amount of communication traffic between the receive antennas and the fusion center.
However, in the LiQuiD-MIMO radar, we propose to use the low-resolution ADCs rather than the high-resolution ADCs such that the system complexity and the data volume that has to be sent to the fusion center will be largely reduced.

Let $\mathcal{Q}_C^{\gamma,b}(\cdot)=\mathcal{Q}^{\gamma,b}(\Re\{\cdot\})+j\mathcal{Q}^{\gamma,b}(\Im\{\cdot\})$ be an element-wise complex-valued quantization operator with $\mathcal{Q}^{\gamma,b}(\cdot)$ given as:
\begin{equation} \label{eqn:quantizer}
	\mathcal{Q}^{\gamma,b}(x)=\left\{\begin{array}{ll}
		-\gamma+\frac{2\gamma}{b}(l+\frac{1}{2}), &\begin{aligned}
			&x-l\frac{2\gamma}{b}+\gamma\in[0,\frac{2\gamma}{b}],\\ &l\in\{0,1,\cdots,b-1\}
		\end{aligned}  \\ \rm{sign}(x)(\gamma-\frac{\gamma}{b}), & \lvert x \rvert > \gamma,
	\end{array}\right.
\end{equation}
where $\gamma$ is the dynamic range of the quantizer, $b$ is the number of quantization levels, i.e., $\tilde{b}=\log_{2}b$.
In practice, $\gamma$ is appropriately chosen to avoid inducing additional distortion due to saturation \cite{Gray1998}. 
With low-resolution ADCs, the value $\tilde{b}$ is much smaller than that with high-resolution ADCs. 
An extreme case is $\tilde{b}=1$, i.e., 1-bit quantization.

Then, in the LiQuiD-MIMO radar, the $\tilde{b}$-bit quantized data is sent to the fusion center.
By considering the induced error during data transmission, the received data at the fusion center can be written as
\begin{equation}
    \label{eqn:Z_mn}
    \mathbf{Z}_{mn}=Q^{\gamma, b}_C(\mathbf{Y}_{mn}) + \tilde{\mathbf{T}}_{mn} = Q^{\gamma, b}_C(\mathbf{X}_{mn}+\mathbf{W}_{mn}) + \tilde{\mathbf{T}}_{mn},
\end{equation}
where $\tilde{\mathbf{T}}_{mn}$ models the data transmission error (DTE).

\vspace{-0.2cm}
\section{Quantized Robust PCA (QRPCA) for LiQuiD-MIMO Radar}
\label{sec:method}
\vspace{-0.1cm}
In this section, we first formulate the problem of recovering the target information matrix (TIM) as a quantized robust PCA (QRPCA) problem, then develop an accelerated proximal gradient (APG)-based algorithm to solve the formulated QRPCA problem.

\vspace{-0.2cm}
\subsection{QRPCA Problem Formulation}
\label{subsec:problem_formulation}
\vspace{-0.1cm}
According to (\ref{eqn:Y_mn}), the TIM matrix $\mathbf{X}_{mn}$ can be decomposed into $\mathbf{X}_{mn}= \mathbf{A}_{mn}\boldsymbol{\Lambda}_{mn}\mathbf{B}_{mn}$. 
It has been shown in \cite{Sun2019TargetEB} that $\textbf{X}_{mn}$ is of low rank, and its rank depends on the number of targets with different distances or different velocities. 
In addition, the DTE $\tilde{\mathbf{T}}_{mn}$ is generally sparse since the bit error rate (BER) is generally quite low.
In addition, the received data $\mathbf{Z}_{mn}$ can be equivalently represented as
\begin{equation}
    \mathbf{Z}_{mn} = Q^{\gamma, b}_C(\mathbf{X}_{mn}+\mathbf{T}_{mn} + \mathbf{W}_{mn}),
\end{equation}
where $\mathbf{T}_{mn}$ is an equivalent sparse DTE before quantization.

For ease of notation, we will omit the subscript $mn$ in the remainder of this section.
By exploiting the classic RPCA formulation, we can simultaneously recover the low-rank matrix $\mathbf{X}$ and the sparse matrix $\mathbf{T}$ by solving the following quantized version of RPCA problem, referred to as the QRPCA problem:
\begin{equation}
\label{prob:QRPCA}
    \min_{ \mathbf{X},\mathbf{T} } \quad \frac{1}{2}D(\mathbf{Z},\mathbf{X}+\mathbf{T}) +\mu\Vert \mathbf{X} \Vert_\ast+\lambda\Vert \mathbf{T} \Vert_1,
\end{equation}
where $D(\cdot,\cdot)$ is some similarity metric which measures the similarity between the quantized data and the unquantized data, and $\mu$ and $\lambda$ denote the regularized parameters.

According to the definition of quantization operator in (\ref{eqn:quantizer}), it can be derived that the quantization data $\mathbf{Z}$ determines the upper bound and the lower bound of the unquantization data $\mathbf{Y}$, i.e.,
\begin{equation}
    \begin{split}
        -\frac{\Delta}{2} \leq \Re\{ \mathbf{Y}-\mathbf{Z} \} \leq \frac{\Delta}{2}\\
        -\frac{\Delta}{2} \leq \Im\{ \mathbf{Y}-\mathbf{Z} \} \leq \frac{\Delta}{2}
    \end{split}
\end{equation}
where $\Delta=\frac{2\gamma}{b}$ denotes the quantization interval. 
By considering the unknown noise distortion on the quantized data, we define the similarity metric function $D(\cdot,\cdot)$ in (\ref{prob:QRPCA}) as follows:
\begin{equation}
\label{penalty}
    \begin{split}
    &D(\mathbf{Z},\mathbf{X}+\mathbf{T})=\\
        &\left\Vert\rho\left( 
        \left[ \Re \{ \mathbf{X}+\mathbf{T}-\mathbf{Z} \} +\frac{\Delta}{2};
        \Im \{ \mathbf{X}+\mathbf{T}-\textbf{Z} \} + \frac{\Delta}{2} \right]
        \right)\right\Vert_F^2
        \\+
        &\left\Vert\rho\left(
        \left[ \Re \{ \mathbf{Z}-\mathbf{X}-\mathbf{T} \} + \frac{\Delta}{2};
        \Im \{ \mathbf{Z}-\mathbf{X}-\mathbf{T} \} + \frac{\Delta}{2} \right]
        \right)\right\Vert_F^2,
    \end{split}
\end{equation}
where $\rho(\cdot)$ is an element-wise function with $\rho(x) = \max\{-x,0\}$.

\vspace{-0.2cm}
\subsection{Accelerated Proximal Gradient Algorithm for QRPCA Problem}
\label{subsec:algorithm}
\vspace{-0.1cm}
Now we introduce an accelerated proximal gradient (APG)-based algorithm \cite{Beck2009} for solving the QRPCA problem (\ref{prob:QRPCA}).
Since the QRPCA problem is convex, the interior point method \cite{cvx} can be used here.
However, it requires high computational complexity and cannot be applied to large-scale problems.
For the real-time implementation of target detection in radar applications, we propose the computationally
efficient APG-QRPCA algorithm.

According to the idea of APG  \cite{Beck2009}, we reformulate the objective function in (\ref{prob:QRPCA}) as
\begin{equation}
\label{pro:APG_QRPCA}
    \min_{ \mathbf{X},\mathbf{T} } \quad h(\mathbf{X},\mathbf{T}) + g(\mathbf{X},\mathbf{T}),
\end{equation}
where $h(\mathbf{X},\mathbf{T})\triangleq \mu\Vert \mathbf{X} \Vert_\ast+\lambda\Vert \mathbf{T} \Vert_1$ and $g(\mathbf{X},\mathbf{T}) \triangleq \frac{1}{2}D(\mathbf{Z},\mathbf{X}+\mathbf{T})$.
During each of the iterations, the algorithm first updates the matrices $\mathbf{X}_l$ and $\mathbf{T}_l$ to $\bar{\mathbf{X}}_l$ and $\bar{\mathbf{T}}_l$, respectively, by carrying some momentum from previous iterations, and then performs the following computation:
\begin{equation}
    \mathbf{X}_p = \bar{\mathbf{X}}_l - \delta \nabla_{\mathbf{X}}g(\bar{\mathbf{X}}_l,\bar{\mathbf{T}}_l),
\end{equation}
\begin{equation}
    \mathbf{T}_p = \bar{\mathbf{T}}_l - \delta \nabla_{\mathbf{T}}g(\bar{\mathbf{X}}_l,\bar{\mathbf{T}}_l),
\end{equation}
\begin{equation}\label{eqn:prox}
    (\mathbf{X}_{l+1},\mathbf{T}_{l+1}) = \text{prox}_{h,\delta}(\mathbf{X}_p,\mathbf{T}_p),
\end{equation}
where $\delta$ is the step size, $\nabla_{\mathbf{X}}g(\mathbf{X},\mathbf{T})$ and $\nabla_{\mathbf{T}}g(\mathbf{X},\mathbf{T})$, respectively, denote the gradients of $g(\mathbf{X},\mathbf{T})$ with respect to $\mathbf{X}$ and $\mathbf{T}$, i.e.,
\begin{equation}
    \begin{split}
       & \nabla_{\mathbf{X}}g(\mathbf{X},\mathbf{T}) = \nabla_{\mathbf{T}}g(\mathbf{X},\mathbf{T})\\
        =&
        \left[\rho\left(\Re(\mathbf{Z}-\mathbf{X}-\mathbf{T})+\frac{\Delta}{2} \right)-
        \rho\left( \Re(\mathbf{X}+\mathbf{T}-\mathbf{Z})+\frac{\Delta}{2} \right)\right] \\
        +j& 
        \left[ \rho\left( \Im(\mathbf{Z}-\mathbf{X}-\mathbf{T})+\frac{\Delta}{2} \right)-
        \rho\left( \Im(\mathbf{X}+\mathbf{T}-\mathbf{Z})+\frac{\Delta}{2} \right) \right].
    \end{split}
\end{equation}
The function $\text{prox}_{h,\delta}(\mathbf{X}_p,\mathbf{T}_p)$ in (\ref{eqn:prox}) denotes the proximal mapping which solves the following problem:
\begin{equation}\label{prob:prox}
\begin{split}
    &\text{prox}_{h,\delta}(\mathbf{X}_p,\mathbf{T}_p)=\\
    &\arg\min_{\mathbf{X},\mathbf{T}} \frac{1}{2\delta}\|\mathbf{X}+\mathbf{T}-\mathbf{X}_p-\mathbf{T}_p\|_F^2 + h(\mathbf{X},\mathbf{T}).
\end{split}
\end{equation}

By solving the problem (\ref{prob:prox}), the proximal mapping can be split into the following two subproblems, which have closed-form expressions:
\begin{equation}
    \mathbf{X}_{l+1}=\arg\min_{\mathbf{X}} \left\{
    \mu \Vert \mathbf{X} \Vert_* + \frac{1}{2\delta} \Vert \mathbf{X}-\mathbf{X}_p\Vert^2_F \right\}
    = \mathbf{U}_p\mathcal{S}_{\mu\delta}(\boldsymbol{\Sigma}_p)\mathbf{V}_p^T
\end{equation}
\begin{equation}
    \textbf{T}_{l+1}=\arg\min_{\mathbf{T}} \left\{
     \lambda \Vert \mathbf{T} \Vert_1 + \frac{1}{2\delta} \Vert \mathbf{T}-\mathbf{T}_p\Vert^2_F \right\}
    = \mathcal{S}_{\lambda\delta}(\mathbf{T}_p)
\end{equation}
where $\mathbf{X}_p=\mathbf{U}_p\boldsymbol{\Sigma}_p\mathbf{V}_p^T$ is the singular value decomposition of $\mathbf{X}_p$, and $\mathcal{S}_{\varepsilon}(\cdot)$ is the element-wise soft-thresholding operator:
\begin{equation}
    \mathcal{S}_{\varepsilon}(x)=\left\{
    \begin{aligned}
    x-\varepsilon, \quad& x>\varepsilon\\
    x+\varepsilon, \quad& x<-\varepsilon\\
    0, \quad& \rm others.
    \end{aligned}
    \right.
\end{equation}
Finally, we can obtain the estimates of the TIM matrix $\mathbf{X}$ and the DTE matrix $\mathbf{T}$, denoted as $\hat{\mathbf{X}}$ and $\hat{\mathbf{T}}$, when the iterations are ended. 

The details of the proposed APG-QRPCA algorithm are given in Algorithm \ref{alg1}.

\begin{algorithm}
    \renewcommand{\algorithmicrequire}{\textbf{Input:}}
    \renewcommand{\algorithmicensure}{\textbf{Output:}}
    \caption{APG-QRPCA Algorithm}
	\label{alg1}
	\begin{algorithmic}[1]
		\REQUIRE $\mathbf{Z}$, $\mu, \lambda, \delta, \text{MaxIter}$ 
            \STATE Initialization: $\mathbf{X}_1 =\mathbf{X}_{0} =0;             \mathbf{T}_1 = \mathbf{T}_{0}=0;\zeta_{0}=1;l=1$
            \WHILE {$l\leq\text{MaxIter}$ or not convergence}
                \STATE $\zeta_l = \frac{1+\sqrt{4\zeta_{l-1}^2+1}}{2}$;
                \STATE $\bar{\mathbf{X}}_l = \mathbf{X}_l + \frac{\zeta_{l-1}-1}{\zeta_l}(\mathbf{X}_l-\mathbf{X}_{l-1}),\bar{\mathbf{T}}_l = \mathbf{T}_l + \frac{\zeta_{l-1}-1}{\zeta_l}(\mathbf{T}_l-\mathbf{T}_{l-1})$;
                \STATE $\mathbf{X}_p = \bar{\mathbf{X}}_l - \delta \nabla_{\mathbf{X}}g(\bar{\mathbf{X}}_l,\bar{\mathbf{T}}_l),  \mathbf{T}_p = \bar{\mathbf{T}}_l - \delta \nabla_{\mathbf{T}}g(\bar{\mathbf{X}}_l,\bar{\mathbf{T}}_l)$;
                \STATE $(\mathbf{U}_p,\boldsymbol{\Sigma}_p,\mathbf{V}_p)=\text{svd}(\mathbf{X}_p),
                \mathbf{X}_{l+1}=\mathbf{U}_p\mathcal{S}_{\mu\delta}(\boldsymbol{\Sigma}_p)\mathbf{V}_p^T,
                \mathbf{T}_{l+1}=\mathcal{S}_{\lambda\delta}(\mathbf{T}_p)$;
                \STATE $l= l+1$
            \ENDWHILE
            \ENSURE $\hat{\mathbf{X}}= \mathbf{X}_l$, $\hat{\mathbf{T}}= \mathbf{T}_l$
	\end{algorithmic}  
\end{algorithm}

\vspace{-0.2cm}
\section{Least Square (LS)-based Target Parameter Estimation}
\label{sec:parameter_estimation}
\vspace{-0.1cm}
By using the proposed APG-QRPCA algorithm, the $M_tM_r$ TIM matrices $\{\mathbf{X}_{mn}, m=1,\cdots,M_t, n=1,\cdots,M_r\}$ can be recovered at the fusion center, which can then be used to estimate the unknown target parameters $\{\mathbf{p}^{(k)},\mathbf{v}^{(k)}\}_{k=1}^K$.
There are a variety of classical signal processing techniques can be used here, including least squares (LS) \cite{Mojtaba2013TAES}, maximum likelihood (ML)\cite{He2010TSP}, as well as sparse reconstruction methods\cite{Gogineni2011TSP}. 
In the following, we introduce a sequential LS method to sequentially estimate the position and velocity parameters.

Let $\boldsymbol{\theta}=\{\boldsymbol{\theta}_p, \boldsymbol{\theta}_v\}$ be the unknown target parameters, where $\boldsymbol{\theta}_p=\{\mathbf{p}^{(k)}\}_{k=1}^K$ and $\boldsymbol{\theta}_v=\{\mathbf{v}^{(k)}\}_{k=1}^K$ denote the position parameters and velocity parameters, respectively. 
According to (\ref{eqn:Y_mn}), $\boldsymbol{\theta}_p$ and $\boldsymbol{\theta}_v$ are implicitly determined by the matrices $\mathbf{A}_{mn}$ and $\mathbf{B}_{mn}$, respectively. 
For the sake of convenience, we use the notations $\mathbf{A}_{mn}(\boldsymbol{\theta}_p)$ and $\mathbf{B}_{mn}(\boldsymbol{\theta}_v)$ in the following discussion.
The target parameter estimation problem can be formulated as the following LS problem:
\begin{equation}\label{prob:LS}
    \hat{\boldsymbol{\theta}} = \arg\min_{\boldsymbol{\theta}} \sum_{m=1}^{M_t}\sum_{n=1}^{M_r}\|\hat{\mathbf{X}}_{mn} - \mathbf{A}_{mn}(\boldsymbol{\theta}_p)\boldsymbol{\Lambda}_{mn} \mathbf{B}_{mn}(\boldsymbol{\theta}_v)\|_2^2.
\end{equation}
which is a computationally demanding problem, requiring searching all possible parameters over high-dimensional space.
Therefore, we separate (\ref{prob:LS}) into two sequential subproblems.

Since the LS solution of the matrix $\boldsymbol{\Lambda}_{mn} \mathbf{B}_{mn}(\boldsymbol{\theta}_v)$ is $(\mathbf{A}_{mn}^H(\boldsymbol{\theta}_p)\mathbf{A}_{mn}(\boldsymbol{\theta}_p))^{-1}\mathbf{A}_{mn}^H(\boldsymbol{\theta}_p)\hat{\mathbf{X}}_{mn}$, we can formulate the following position estimation problem:
\begin{equation}\label{prob:LS_position}
    \hat{\boldsymbol{\theta}}_p = \arg\min_{\boldsymbol{\theta}_p} \sum_{m=1}^{M_t}\sum_{n=1}^{M_r}\|\mathbf{P}_{p,mn}^\perp(\boldsymbol{\theta}_p)\hat{\mathbf{X}}_{mn}\|_2^2,
\end{equation}
where $\mathbf{P}_{p,mn}^\perp(\boldsymbol{\theta}_p) = \mathbf{I} - \mathbf{A}_{mn}(\boldsymbol{\theta}_p)(\mathbf{A}_{mn}^H(\boldsymbol{\theta}_p)\mathbf{A}_{mn}(\boldsymbol{\theta}_p))^{-1}\mathbf{A}_{mn}^H(\boldsymbol{\theta}_p)$.
The above problem can be solved by searching over the two-dimensional plane where the targets are located.

Once the position parameters are estimated, we also get an estimate of the matrix $\boldsymbol{\Lambda}_{mn} \mathbf{B}_{mn}(\boldsymbol{\theta}_v)$, denoted as $\hat{\boldsymbol{\Xi}}_{mn}$. 
Then, by performing FFT over each row of $\hat{\boldsymbol{\Xi}}_{mn}$, the Doppler frequency $f_{mn}^{(k)}$ of each target corresponding to each pair of Tx-Rx antennas can be estimated, denoted as $\hat{f}_{mn}^{(k)}$.
Then the velocity parameters can then be estimated by solving the following LS problem
\begin{equation}\label{prob:LS_velocity}
    \hat{\boldsymbol{\theta}}_v = \arg\min_{\boldsymbol{\theta}_v} \sum_{m=1}^{M_t}\sum_{n=1}^{M_r}\sum_{k=1}^K (\hat{f}_{mn}^{(k)}-f_{mn}^{(k)}(\mathbf{v}_k))^2,
\end{equation}
where the function $f_{mn}^{(k)}(\mathbf{v}_k)$ is given by (\ref{eqn:f_mn}).

\begin{figure}[t]
\centering
\includegraphics[width=\figWidth, height=\figHeight]{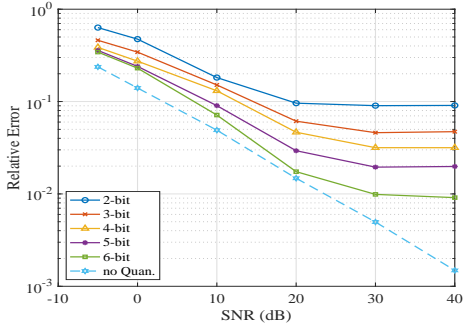}
\vspace{-0.1cm}
\caption{Performance of recovering the low-rank matrix $\mathbf{X}_{mn}$ versus different SNRs.}
\label{fig:lowrank_matrix}
\end{figure}

\vspace{-0.2cm}
\section{Numerical Results}
\label{sec:Sims}
\vspace{-0.1cm}
Now we evaluate the performance of our proposed algorithm for the LiQuiD-MIMO radar system.
Throughout the simulations, we consider a MIMO radar equipped with $M_t=3$ transmit antennas and $M_r=10$ receive antennas uniformly distributed on the concentric circles with radius $5$km and $3$km, respectively. 
The reference carrier frequency parameters $f_0=5$GHz and the frequency increment $\Delta f=50$MHz. One CPI consists of $Q=128$ pulses with $T_{\rm{PRI}}=0.5$ms and $T_p=6.4 \mu \rm{s}$.
The Hadamard sequence is used as the baseband waveform with $B_0 =10$MHz.
At the fusion center, $1\%$ of the received data is assumed to be received incorrectly, leading to a sparse DTE matrix.
We consider the scenario that one target is located at $\mathbf{p}^{(1)}=[1100, 1100]^T$m with $\mathbf{v}^{(1)}=[10,10]^T$ m/s.
The performance of the classic RPCA algorithm without considering the effect of quantization is also demonstrated as a benchmark, denoted as ``No Quan.'' method.

Fig.1 and Fig.2 depict the relative errors in recovering the low-rank matrix $\mathbf{X}_{mn}$ and the sparse matrix $\mathbf{T}_{mn}$, respectively, with respect to different SNR values, by relying on the proposed APG-QRPCA algorithm.
It is shown that it is possible to simultaneously recover the  matrices $\mathbf{X}_{mn}$ and $\mathbf{T}_{mn}$ from the low-bit quantized data.
When the SNR is less than $20$dB, the performance of 6-bit quantization is very close to that without quantization, proving the effectiveness of low-bit quantization.
However, there exists a performance gap between the low-bit quantization and that without quantization when SNR is higher than $20$dB.

\begin{figure}[t]
\centering
\includegraphics[width=\figWidth, height=\figHeight]{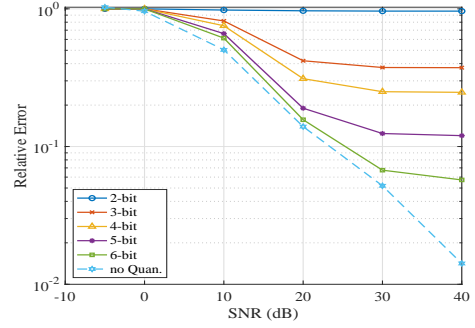}
\vspace{-0.1cm}
\caption{Performance of recovering the sparse matrix $\mathbf{T}_{mn}$ versus different SNRs.}
\label{fig:sparse_matrix}
\end{figure}

\begin{figure}
   \centering
   \begin{subfigure}[b]{0.48\linewidth}
      \centering
      \includegraphics[width=\linewidth]{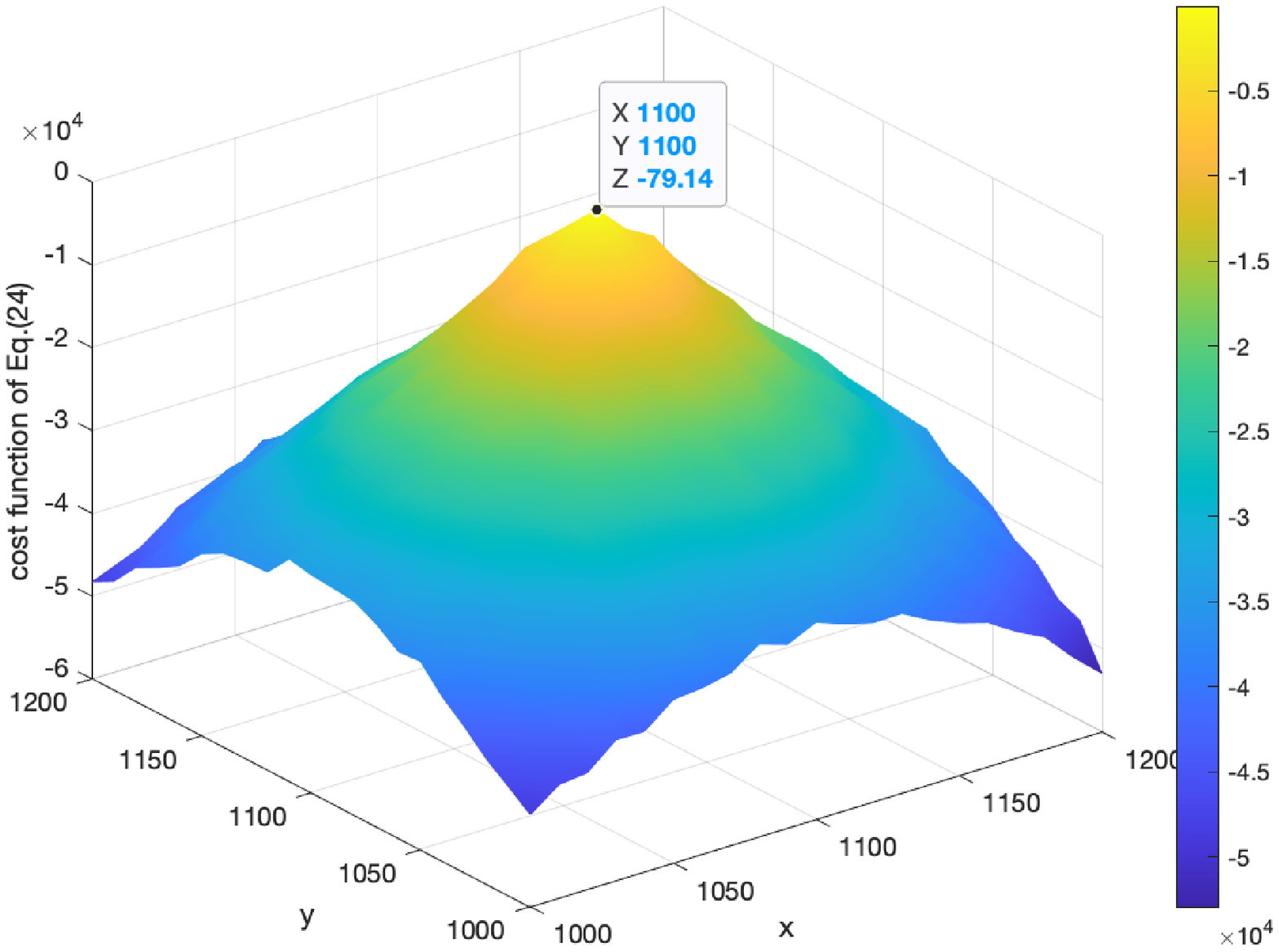}
      \caption{Position estimation}
      \label{fig:lowrank_matrix}
    \end{subfigure} 
    \hfill
    \begin{subfigure}[b]{0.48\linewidth}
      \centering
      \includegraphics[width=\linewidth]{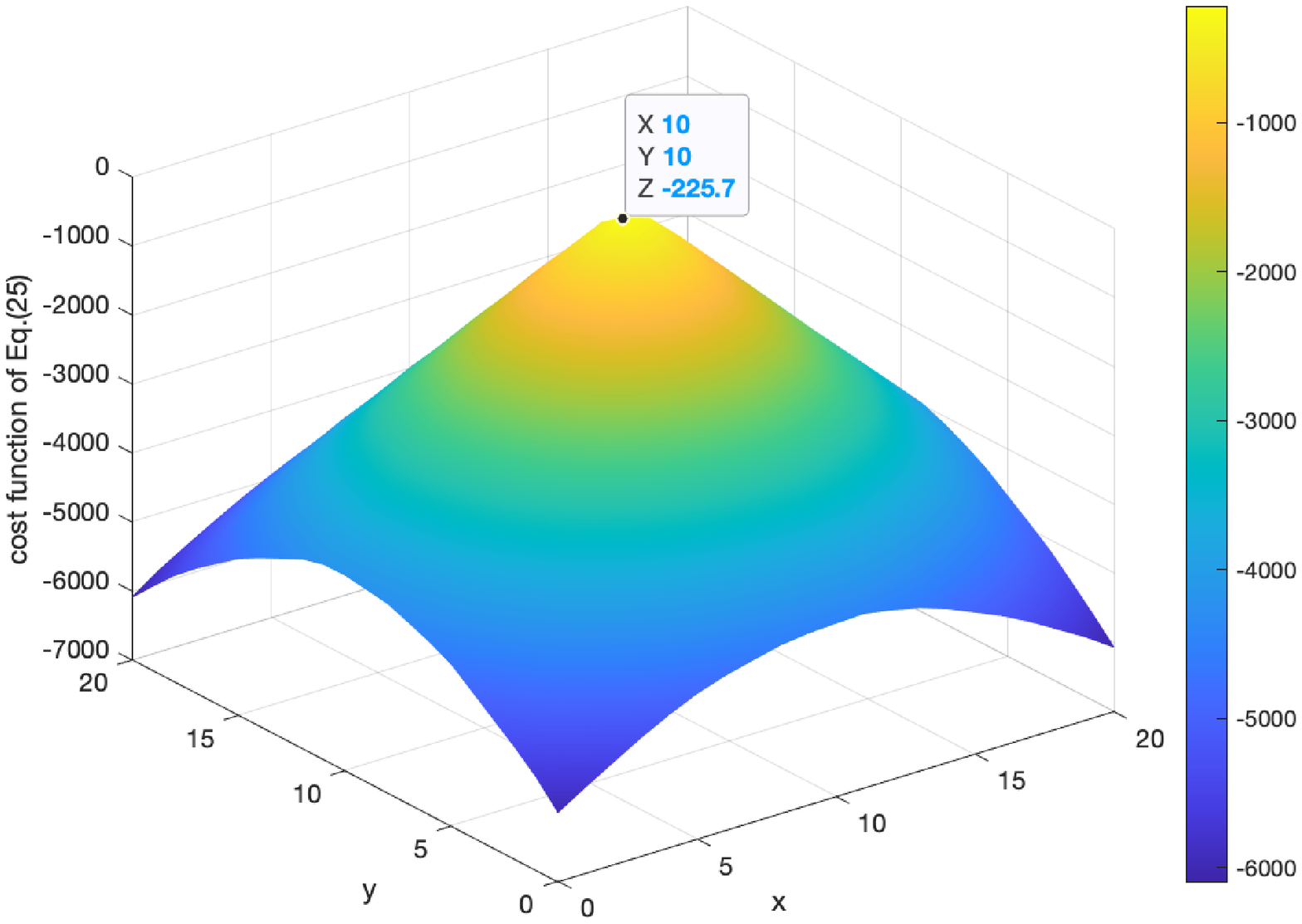}
      \caption{Velocity estimation.}
      \label{fig:sparse_matrix}
    \end{subfigure} 
    \caption{The results of the target parameter estimation for a target at $\mathbf{p}=[1100, 1100]^T$m with velocity $\mathbf{v}=[10,10]^T$m/s, by using $4$-bit quantization at $\text{SNR}=20$dB.} 
    \label{fig:QRPCA} 
\end{figure}

After recovering the set of matrices $\{\mathbf{X}_{mn}, m=1,\cdots, M_t, n=1,\cdots, M_r\}$, we now show the performance of target parameter estimation using the LS-based method proposed in Section \ref{sec:parameter_estimation}.
Fig.3 shows the results of position estimation and velocity estimation at $\text{SNR}=20$dB by solving the problems (\ref{prob:LS_position}) and (\ref{prob:LS_velocity}), respectively.
It can be seen that the position and velocity of the target can be accurately estimated by using the LS-based method.

\vspace{-0.2cm}
\section{Conclusion}
\label{sec:conclusion}
\vspace{-0.1cm}
In this work we proposed a low-bit quantized distributed MIMO radar system, which significantly reduces the system complexity of the widely-separated receivers and the data transmission volume between the receivers and the fusion center.
The main challenge in such LiQuiD-MIMO radar is to deal with the quantization distortion induced by low-resolution ADCs. 
By exploiting the low-rank structure of the target information matrix and the sparsity of the data transmission error, we formulated a quantized version of RPCA problem, referred to as QRPCA problem, and developed an APG-based algorithm, refered to as APG-QRPCA algorithm, to simultaneously recover the infinite-precision target information matrix and the errors during data transmission. 
Simulation results demonstrate that it is feasible to implement a distributed MIMO radar system by employing low-resolution ADCs.
The analysis of the performance bound of the proposed LiQuiD-MIMO radar will be our future work.

\bibliographystyle{IEEEtran}
\bibliography{IEEEabrv,reference}

\end{document}